\documentclass{aastex63}

\usepackage{CJK}

\received{}
\revised{}
\accepted{}
\submitjournal{}

\shorttitle{The ANTARES Event Broker}
\shortauthors{Matheson et al.}

\begin{document}
\begin{CJK*}{UTF8}{bkai}

\title{The ANTARES Astronomical Time-Domain Event Broker}

\correspondingauthor{Thomas Matheson}
\email{tom.matheson@noirlab.edu}

\author[0000-0001-6685-0479]{Thomas Matheson}
\affiliation{NSF's National Optical-Infrared Astronomy Research Laboratory\\
950 North Cherry Avenue\\
Tucson, AZ 85719, USA}

\author[0000-0002-2744-714X]{Carl Stubens}
\affiliation{NSF's National Optical-Infrared Astronomy Research Laboratory\\
950 North Cherry Avenue\\
Tucson, AZ 85719, USA}

\author[0000-0002-9508-1629]{Nicholas Wolf}
\affiliation{NSF's National Optical-Infrared Astronomy Research Laboratory\\
950 North Cherry Avenue\\
Tucson, AZ 85719, USA}

\author[0000-0003-1700-5740]{Chien-Hsiu Lee (李見修)}
\affiliation{NSF's National Optical-Infrared Astronomy Research Laboratory\\
950 North Cherry Avenue\\
Tucson, AZ 85719, USA}

\author[0000-0001-6022-0484]{Gautham Narayan}
\affiliation{Department of Astronomy \\
University of Illinois at Urbana-Champaign \\
Urbana, IL 61801, USA}
\affiliation{Center for Astrophysical Surveys, National Center for Supercomputing Applications \\ Urbana, IL 61801, USA}

\author[0000-0002-6839-4881]{Abhijit Saha}
\affiliation{NSF's National Optical-Infrared Astronomy Research Laboratory\\
950 North Cherry Avenue\\
Tucson, AZ 85719, USA}

\author[0000-0002-1140-5463]{Adam Scott}
\affiliation{NSF's National Optical-Infrared Astronomy Research Laboratory\\
950 North Cherry Avenue\\
Tucson, AZ 85719, USA}

\author{Monika Soraisam}
\affiliation{National Center for Supercomputing Applications \\
University of Illinois at Urbana-Champaign \\
Urbana, IL 61801, USA}
\affiliation{Department of Astronomy \\
University of Illinois at Urbana-Champaign \\
Urbana, IL 61801, USA}

\author[0000-0002-9836-603X]{Adam S.Bolton}
\affiliation{NSF's National Optical-Infrared Astronomy Research Laboratory\\
950 North Cherry Avenue\\
Tucson, AZ 85719, USA}

\author{Benjamin Hauger}
\affiliation{NSF's National Optical-Infrared Astronomy Research Laboratory\\
950 North Cherry Avenue\\
Tucson, AZ 85719, USA}

\author[0000-0002-7678-2155]{David R.Silva}
\affiliation{University of Texas at San Antonio College of Sciences \\
One UTSA Circle \\
San Antonio, TX 78249, USA}

\author[0000-0003-1204-6535]{John Kececioglu}
\affiliation{Department of Computer Science \\
The University of Arizona \\
1040 East 4th Street \\
Tucson, AZ 85721, USA}

\author{Carlos Scheidegger}
\affiliation{Department of Computer Science \\
The University of Arizona \\
1040 East 4th Street \\
Tucson, AZ 85721, USA}

\author[0000-0003-4703-7460]{Richard Snodgrass}
\affiliation{Department of Computer Science \\
The University of Arizona \\
1040 East 4th Street \\
Tucson, AZ 85721, USA}

\author[0000-0002-6298-1663]{Patrick D. Aleo}
\affiliation{Department of Astronomy \\
University of Illinois at Urbana-Champaign \\
Urbana, IL 61801, USA}

\author{Eric Evans-Jacquez}
\affiliation{Department of Computer Science \\
The University of Arizona \\
1040 East 4th Street \\
Tucson, AZ 85721, USA}
\affiliation{Wisconsin IceCube Particle Astrophysics Center \\
University of Wisconsin-Madison \\
Madison, WI 53703 USA}

\author{Navdeep Singh}
\affiliation{Department of Computer Science \\
The University of Arizona \\
1040 East 4th Street \\
Tucson, AZ 85721, USA}
\affiliation{Amazon.com, Inc \\
440 Terry Ave N \\
Seattle, WA 98109, USA}

\author{Zhe Wang}
\affiliation{Department of Computer Science \\
The University of Arizona \\
1040 East 4th Street \\
Tucson, AZ 85721, USA}

\author{Shuo Yang}
\affiliation{Department of Computer Science \\
The University of Arizona \\
1040 East 4th Street \\
Tucson, AZ 85721, USA}
\affiliation{DiDi Research America \\
450 National Avenue \\
Mountain View, CA 94043, USA}

\author{Zhenge Zhao}
\affiliation{Department of Computer Science \\
The University of Arizona \\
1040 East 4th Street \\
Tucson, AZ 85721, USA}




\begin{abstract}

We describe the Arizona-NOIRLab Temporal Analysis and Response to Events System (ANTARES), a software instrument designed to process large-scale streams of astronomical time-domain alerts.  With the advent of large-format CCDs on wide-field imaging telescopes, time-domain surveys now routinely discover tens of thousands of new events each night, more than can be evaluated by astronomers alone.  The ANTARES event broker will process alerts, annotating them with catalog associations and filtering them to distinguish customizable subsets of events.  We describe the data model of the system, the overall architecture, annotation, implementation of filters, system outputs, provenance tracking, system performance, and the user interface.

\end{abstract}



\section{Introduction} \label{sec:intro}

The practice of time-domain astronomy is undergoing immense changes.
The pathway for detection of transient and variable objects in the sky
once relied on visual inspection and comparison of images.  The advent
of digital detectors, especially large-format CCDs that can be
deployed on wide-field telescopes, has enabled a new era where image
subtraction algorithms can identify virtually every difference
between a current image and a template over thousands of square degrees of sky each night.  

As an illustration of the change in scale, there were 4038 International Astronomical Union Circulars (IAUCs) issued between 1991 January and 2010 December with 10,358 entries.  Some of those entries were for multiple objects, but the vast majority were single events, sometimes for follow-up observations of prior discoveries.  Even if we make the generous assumption of $\sim$2 unique objects per IAUC entry, the main world-wide channel for information about transients and variables had $\sim$20,000 events over two decades, averaging fewer than three per day.  The Zwicky Transient Facility \citep[ZTF,][]{bellm19, graham19, masci19} uses a 47 square degree imager on the Palomar 48-inch telescope to perform a time-domain survey.  The public portion of the ZTF survey (40\% of the time) generates several hundred thousand alerts (any object that has changed significantly in brightness or position) each night and thus outperforms those two recent decades in less than an hour.

On the not-too-distant horizon is the Legacy Survey of Space and Time
\citep[LSST,][]{ivezic08, kantor14, kahn18} that will be carried out at 
the Vera C. Rubin Observatory.  With a 10 square degree camera and 
8.4m mirror (effective aperture 6.7m), the Rubin Observatory will be
able to observe a wide, fast, and deep survey of the southern sky.
Each LSST image is anticipated to have $\sim$10,000 objects that have
changed in brightness or position at a significance of 5$\sigma$.
With $\sim$1000 pointings per night, LSST will generate
$\sim$10,000,000 time-domain alerts every night for the ten-year
duration of the survey.  Many of these will be repetitions of periodic
variables or known Solar System objects \citep[e.g.,][]{ridgway14}, but even
10\% of the LSST rate will be enormous.  Hidden among these unknown
events will be rare and unusual objects, some never seen before and
some with short lifetimes.  Without an efficient software
infrastructure to filter these alerts, the events that generated them will remain undiscovered.
Building such a system has been a high priority for national
committees concerned with maximizing the science potential of LSST
\citep[e.g., ][]{elmegreen15, najita16} and this led to the development of a software infrastructure system
that can process a large volume of alerts efficiently and rapidly.  In
2013, astronomers at the National Optical Astronomy Observatory (NOAO) and computer scientists from the University
of Arizona began to create this tool, now named the Arizona-NOIRLab
Temporal Analysis and Response to Events System (ANTARES).  

A system that acts as an intermediary between sources of information
and the consumer in the commercial world has come to be known as a
broker, with a stock broker being perhaps the most common example.
This term has been adopted by the astronomical community
\citep[e.g.,][]{borne08}, so ANTARES and systems like it are often
referred to as alert brokers (or, simply, brokers).  For a summary of
prior work in brokering, see Section 2.3 of \citet{narayan18}.  Other
broker systems include Make Alerts Really Simple
(MARS),\footnote{\url{https://mars.lco.global/}} Lasair
\citep{smith19}, Alert Management, Photometry and Evaluation of
Lightcurves, \citep[AMPEL,][]{nordin19}, Automatic Learning for the
Rapid Classification of Events \citep[ALeRCE,][]{forster20}, and Fink
\citep{moller20}.  Alert brokers can filter alerts, add value through
annotation from many sources of information, characterize and classify
alerts, and distribute alerts to the consumer.

Overall, we view ANTARES as an astronomical instrument, but one that lets
scientists collect and analyze time-domain alerts, rather than
photons.  ANTARES is designed to operate at LSST scale and beyond while ingesting
alerts, annotating them with additional information, filtering them,
and distributing alerts of interest to astronomers who request them.
In this paper, we describe the principles that guided development, the
overall architecture of the system, how the various components work,
and the performance as deployed at the ZTF scale.  Descriptions of previous versions of the system were presented by \citet{saha14, saha16} and \citet{narayan18}.

\section{Illustrative Use Cases}

The overarching design goal for ANTARES is to enable the broadest
possible scientific use of time-domain alerts generated by large-scale
surveys.  ANTARES provides a flexible infrastructure that allows the
end user to filter and classify alerts in virtually any way that is of
interest to them.  The scientific aims of ANTARES will be determined
largely by the demands and interests of the community that uses it.
An exhaustive list of all possible uses is beyond the scope of this
paper, but we can describe several use cases that illustrate the
requirements for the system.  These cases will demonstrate the logic
behind the capabilities of ANTARES.

Time-domain surveys with high cadence have enabled detections of transients shortly after they occur.  A prime example is SN~2011fe, a Type Ia supernova (SN) in M101, discovered by the Palomar Transient Facility \citep[PTF, the precursor to ZTF,][]{law09} just eleven hours after explosion \citep{nugent11}.  This rapid identification put constraints on the size of the progenitor, demonstrating that the SN arose from a compact object \citep{bloom12snfe}.  Other cases of observations soon after explosions of SNe have revealed the nature of nearby circumstellar material, and thus the evolutionary state of the progenitor star \citep[e.g.,][]{khazov16, dejaeger18, kochanek19, gangopadhyay20, goldberg20, chugai20}.  In general, astronomers will prefer as complete a picture as possible of the evolution of any transient, so early detection is intrinsically valuable.  An effective brokering system will provide early detection and recognition.  To do so, it must process time-domain alerts in real-time, with latency as low as possible.  The LSST alert production system aims to deliver raw alerts within 60 seconds of the shutter closing \citep{kantor14}.  In addition, the brokering system must distinguish interesting transients in the alert stream, so it must have a filtering system that uses the alert data, preferably together with any available external sources of information, to automatically flag candidates.  The filter could even compare alert data with predicted features from models of objects as yet unobserved.

Identification of some transients benefits greatly from external information.  Tidal disruption events (TDEs) can shed light on the nature of black holes, including accretion mechanisms and jet formation \citep{rees88, metzger16, decolle19, hung20, Lee20TDE}.  While the number of discovered TDEs is growing \citep[e.g.,][]{komossa15, hung18, vanvelzen20}, the field is still relatively young.  TDEs tend to occur in post-starburst galaxies \citep{french16, french17, cen20}, although the physical reasons for this are, as yet, unknown.  Catalogs of likely TDE hosts exist \citep{french18}, so a brokering system can use that knowledge as prior information in a filter for TDE detection.  In addition, as TDEs occur at the centers of galaxies, they can be confused with active galactic nuclei (AGN).  A broker can identify AGN either through catalogs of known objects, but also from historical detections at radio or x-ray (or other high-energy) wavelengths.  A broker must then be flexible in incorporating and updating catalogs of known astronomical objects, especially across the full range of the electromagnetic spectrum to use as much available data as possible in filtering alerts.

As described above, the anticipated number of nightly events that LSST will produce for relatively common transients and variables is quite large \citep[e.g.,][]{ridgway14}.  It may not be necessary to obtain spectroscopic and photometric follow up of every one of these transients and variables, but given the limited resources \citep[e.g.,][]{kulkarni20} and the number of such transients, that scale of follow up is not even possible.  Nonetheless, astronomers will still want to follow and characterize subsets of these objects, perhaps in concert with other observing campaigns.  An example could be a program to use Type Ia SNe as cosmological probes (see, e.g., the LSST Dark Energy Science Collaboration Science Road Map\footnote{\url{https://doi.org/10.5281/zenodo.3588457}
}).  The program may have dedicated spectroscopic resources for only a short time, so a steady stream of Type Ia SNe is only needed for that period.  Different groups will have different needs at different times.  The broker must be capable of filtering alerts to generate streams of objects and flexible enough to provide these customized streams on demand.

Rare events detected by time-domain surveys will also be of great interest.  The recent success in identifying an electromagnetic counterpart to a gravitational wave detection \citep[e.g.,][]{abbott17NSdiscover, abbott17NSfollow} highlights the scientific value in distinguishing the rare events from the many more common ones that will turn up in the same survey.  Many multi-messenger events, whether gravitational waves \citep[e.g., LIGO/Virgo,][]{aasi15, acernese15}, neutrinos \citep[e.g.,][]{adrianmartinez11, aartsen17search}, or cosmic rays \citep[e.g.,][]{aharonian06, holder06, knodlseder20}, have poor sky localizations \citep[e.g.,][]{abbott18}, necessitating wide-area searches that will produce large numbers of alerts ($\sim$1000 per square degree, at LSST depth) in a short time.  It may be possible to use the multi-messenger data to identify potential host galaxies and restrict the search area \citep{gehrels16}, but there will still be a large number of events to sort.  In addition, as with other transients, early detection will be valuable \citep[e.g.,][]{arcavi18}.  Even without a multi-messenger component, rare events will appear in time-domain surveys.  At LSST rate, there could be several one-in-a-billion events every year.  The broker must operate rapidly to filter the alerts and distinguish rare events.  One way of finding the needle in the haystack is to remove the hay, so the broker should be able to identify common transients and variables that would otherwise obscure rare events with their numbers.  As an example, flares on dwarf stars could be distinguished using multiwavelength data such as checking for coincidence with red sources in the \emph{WISE} catalog \citep[e.g.,][]{davenport12}.

Activity on Solar System objects provides another tool for investigating their nature and thus elucidating aspects of Solar System formation \citep[e.g.,][]{jewitt15}.  Identification of new moving objects in time-domain surveys employs a distinctly different approach than finding objects that merely change in brightness.  It typically requires combining tracks of objects over multiple frames separated by hours to days \citep[e.g.,][]{trilling17, jones18}.  Finding new moving objects is thus beyond the capacity of a broker such as ANTARES, and it is a task that will be a component of LSST data management \citep{juric17}. Known Solar System objects, however, can be distinguished if the ephemerides are correct.  Both ZTF and LSST will flag known objects and these can then be evaluated to see if they have the expected brightness or if they are showing signs of activity.  Such a specialized process on a large subset of a time-domain stream would likely require a separate broker, so a general-purpose broker must be able to efficiently redirect a significant portion of the alert stream to downstream, customized brokers.

Many variable objects do not need immediate spectroscopic or photometric follow up, especially those with relatively long time scales.  These objects, whether periodic or aperiodic, are most valuable to an astronomer once a full light curve is available.  Depending on the cadence of the survey and the period of the variable, this may take many periods to sufficiently sample the light curve.  To study these objects, real-time filtering is less useful than easy discovery in an archive of alerts.  While surveys may retain searchable archives of raw alerts, the broker must keep a record of the full, annotated set of alerts.  This value-added archive must be built up by the broker over the course of the survey.  It must provide a database of alerts with a flexible query system that enables science on longer time-scale objects.

Sometimes variables (and some longer-lived transients) can behave in unexpected ways.  Examples could include a known pulsating periodic variable suddenly undergoing an eclipse from a distant companion or an eruption of a luminous blue variable.  Wide-field time-domain surveys provide astronomers with a tool that monitors all of these objects.  The broker must provide a way for users to provide watch lists of objects of interest and a method of communication to notify these users when an alert occurs for one of their objects.  

Finally, as each new time-domain survey has begun, the novel aspects of parameter space that it reaches have inevitably yielded interesting objects that no one had yet detected such as superluminous supernovae \citep[e.g.,][]{quimby07, galyam12}, luminous red novae \citep[e.g.,][]{martini99, kasliwal11}, calcium-rich transients \citep[e.g.,][]{kasliwal12ca, foley15}, .Ia supernovae \citep[e.g.,][]{kasliwal10, perets10}, and fast, blue optical transients \citep[e.g.,][]{drout14, smartt18, margutti19}.  Astronomy has traditionally been a science of discovery and exploration.  Wide-field time-domain surveys continue this tradition, providing opportunities for discovery.  There are, as yet, still empty regions in the energy-time diagram of \cite{kasliwal12}.  LSST will produce time-domain alerts at an unprecedented rate, but also at an unprecedented depth, relatively high cadence, and in multiple filters, thus opening an entirely new discovery volume. We may not know precisely what will appear in the LSST alert stream, but we can predict that new and extraordinary objects will be there.  We need to have a broker that can sort through the alerts rapidly, filter them to recognize what we know, and then provide a substream of unusual events that will provide the next breakthrough in astronomy.

\section{Data Model}\label{sec:datamodel}

\begin{figure}
    \centering
    \includegraphics[width=0.7\textwidth]{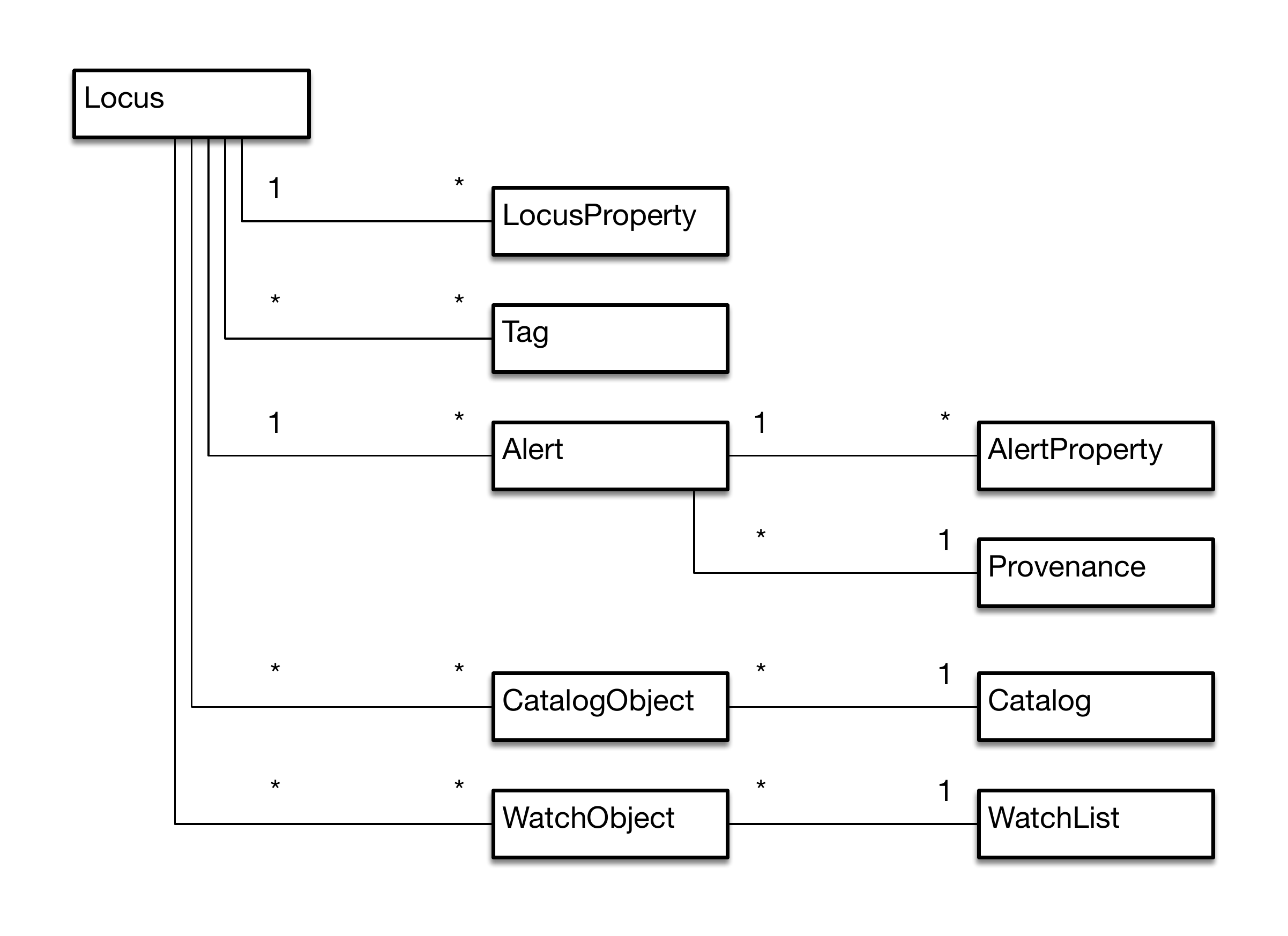}
    \caption{ANTARES data model, simplified. Boxes indicate entities. Lines indicate relationships. Asterisks and numbers on lines indicate relationship type: one-to-many (1---*), many-to-one (*---1), or many-to-many (*---*). Data contained within entities are omitted. The Filter is a software entity, not a data entity, and is also omitted.}
    \label{fig:datamodel}
\end{figure}

A data model is a set of data entities that exist within a system, the relationships between entities, and the way of representing these entities and relationships in a database. The ANTARES data model is driven by the nature of time-domain data. The primary data entities in ANTARES are the Alert and the Locus. Secondary entities include the Tag, LocusProperty, AlertProperty, WatchList, WatchObject, Catalog, and CatalogObject. The data model is visualized in Figure \ref{fig:datamodel}.

An Alert is a message received by ANTARES from a survey such as ZTF or LSST, plus additional properties generated by ANTARES. Typically an Alert is either an observed magnitude measurement or an ``upper limit'' magnitude measurement, but in principle an Alert can be any packet of data that has precise spatial coordinates and a timestamp.\footnote{Region-based alerts (e.g., gravitational wave detections from LIGO/Virgo) do not have definitive coordinate locations, and are handled independently of the Alert/Locus concept. We have implemented a system for ingesting LIGO/Virgo alerts that can be generalized to any alert source with imprecise localizations (and thus large areas associated with the alert), including the neutrino and cosmic ray observatories mentioned above, but also for networks that collate multi-messenger alerts like the Astrophysical Multimessenger Observatory Network \citep[AMON,][]{ayala20}. Filters in ANTARES have access to all LIGO/Virgo data within the past 30 days. This makes it possible to write a filter that flags all Alerts within a particular confidence level region of a recent LIGO/Virgo detection. By combining this with other criteria, such as galaxy catalog associations or light curve history, this allows the production of filters that identify likely LIGO/Virgo correlates.}

A Locus (plural, Loci) is a point on the sky where Alerts cluster and is roughly equivalent to an astrophysical object. The association of Alerts to Loci uses a 1\arcsec\ radius cone search. That is, each incoming Alert is associated with the nearest Locus within 1\arcsec\ of the Alert. If no such Locus exists, a new Locus is created.

Tags are short strings that are associated with one or more Loci. Tags are used to flag loci that meet specific criteria, e.g., ``extragalactic,'' ``nuclear\_transient,'' etc. Tags are used to generate output streams and are queryable in the database.

Alert properties are associated with a particular Alert, such as magnitude. Some originate from the incoming source alert, and others are annotations generated by Filters.

Locus properties are associated with a Locus (i.e., an astronomical object), not an Alert. Examples include color, or statistics of a light curve such as skewness.

CatalogObjects are entities in astronomical object catalogs. ANTARES 1.0 supports extended objects by modeling them as circular regions. Support for other shapes is possible in future versions.

WatchObjects and WatchLists allow users to define and track objects and regions of interest to them. A WatchObject is a coordinate and a radius, and a WatchList is a collection of such objects. Users may create their own WatchLists using the ANTARES Portal (Section \ref{sec:portal} describes the portal) and browse matching Loci. ANTARES can notify users of new hits to their WatchLists over Slack.\footnote{\url{https://slack.com/}}

The Provenance object stores information about the version and configuration of ANTARES that processes each Alert.  It is described further in Section \ref{sec:provenance}.

\section{System Architecture}

The ANTARES system is designed around a parallel stream-processing architecture.  The system ingests alerts from input Apache Kafka streams \citep[Kafka is an industry standard technology for data streaming,][]{kreps11}\footnote{\url{https://kafka.apache.org/}}, processes them, stores them in a database, and produces streams of output alerts using Kafka. The system is composed of ten subsystems that perform specific functions. Figure \ref{fig:layercake} depicts the layering of systems and activities. Figure \ref{fig:architecture} depicts the architecture of the system and connections between components. Figure \ref{fig:pipeline} depicts the behavior of the Alert Pipeline subsystem that processes alert streams. The remainder of this section describes these concepts in more detail.

\subsection{System Layers}

\begin{figure}[p]
    \vspace*{1cm}
    \makebox[\linewidth]{
        \includegraphics[width=0.82\linewidth]{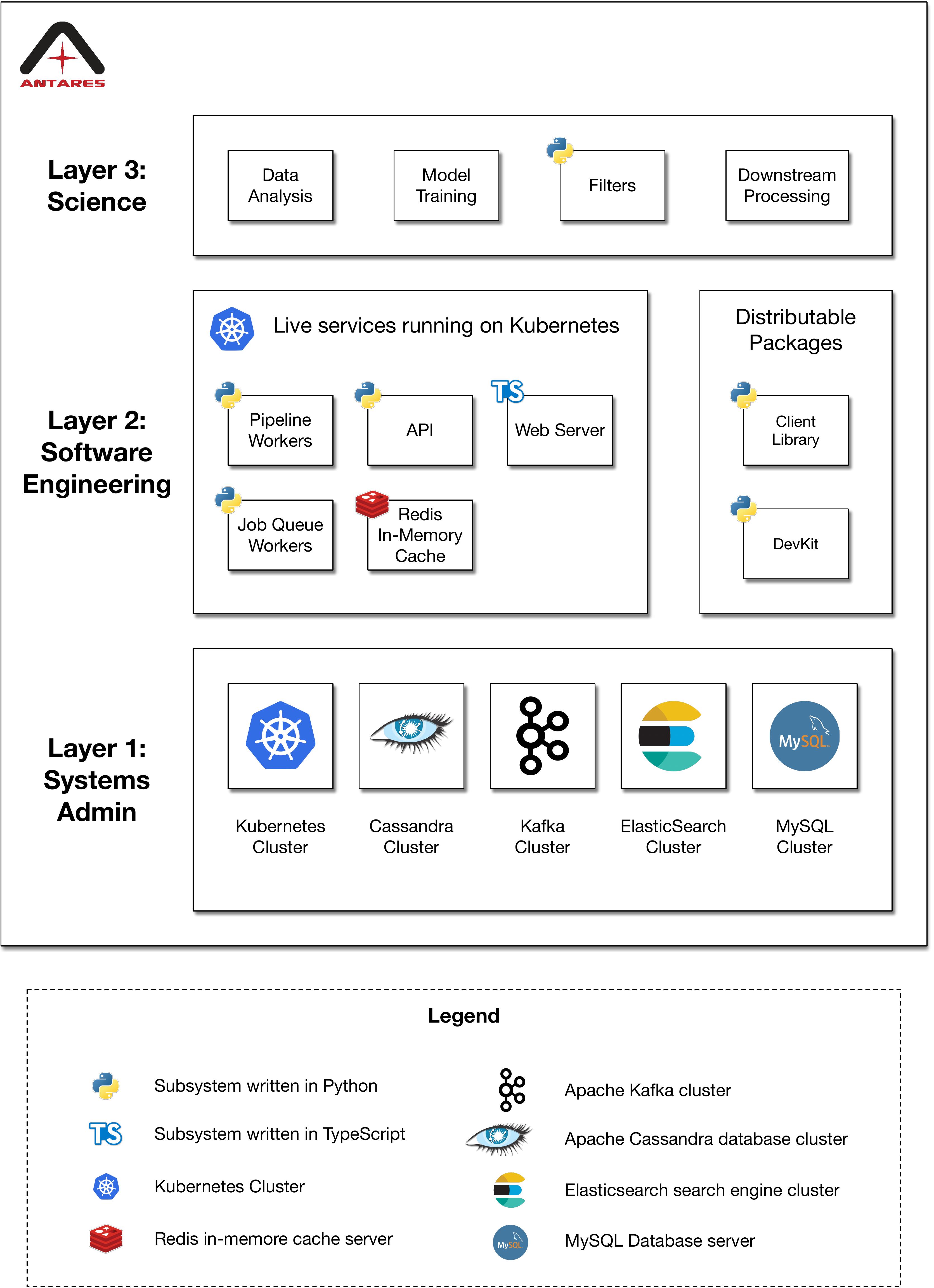}
    }
    \caption{``Layercake'' diagram showing the layers of systems and activities in ANTARES. The Systems Administration layer provides compute and database services. The Software Engineering layer provides software systems. The Science layer consists of activities performed by science staff and researchers from the community. This is an abstract representation, not a system architecture diagram.}
    \label{fig:layercake}
\end{figure}

At an abstract level, the system is composed of three layers as shown in Figure \ref{fig:layercake}. From bottom to top these layers are systems administration, software engineering, and science. Each layer contains responsibilities and systems that facilitate those of the layer above. The systems administration layer is responsible for hosting and managing components that run on bare-metal hardware. This includes the databases, the Kafka streaming message broker, and the Kubernetes\footnote{\url{https://kubernetes.io/}} (a software container orchestration system) compute cluster. On top of this layer is the software engineering layer, which includes two types of systems, live systems that run on Kubernetes and distributable software packages that users download and run outside the ANTARES cluster. On top of the software engineering layer is the science layer. The science layer consists of tasks and components written by researchers who use ANTARES as a science tool. This includes data analysis tasks, model training, filter development, and downstream processing of ANTARES output streams.

\subsection{System Components}

\begin{figure}[p]
    \vspace*{0.1cm}
    \makebox[\linewidth]{
        \includegraphics[width=1.05\linewidth]{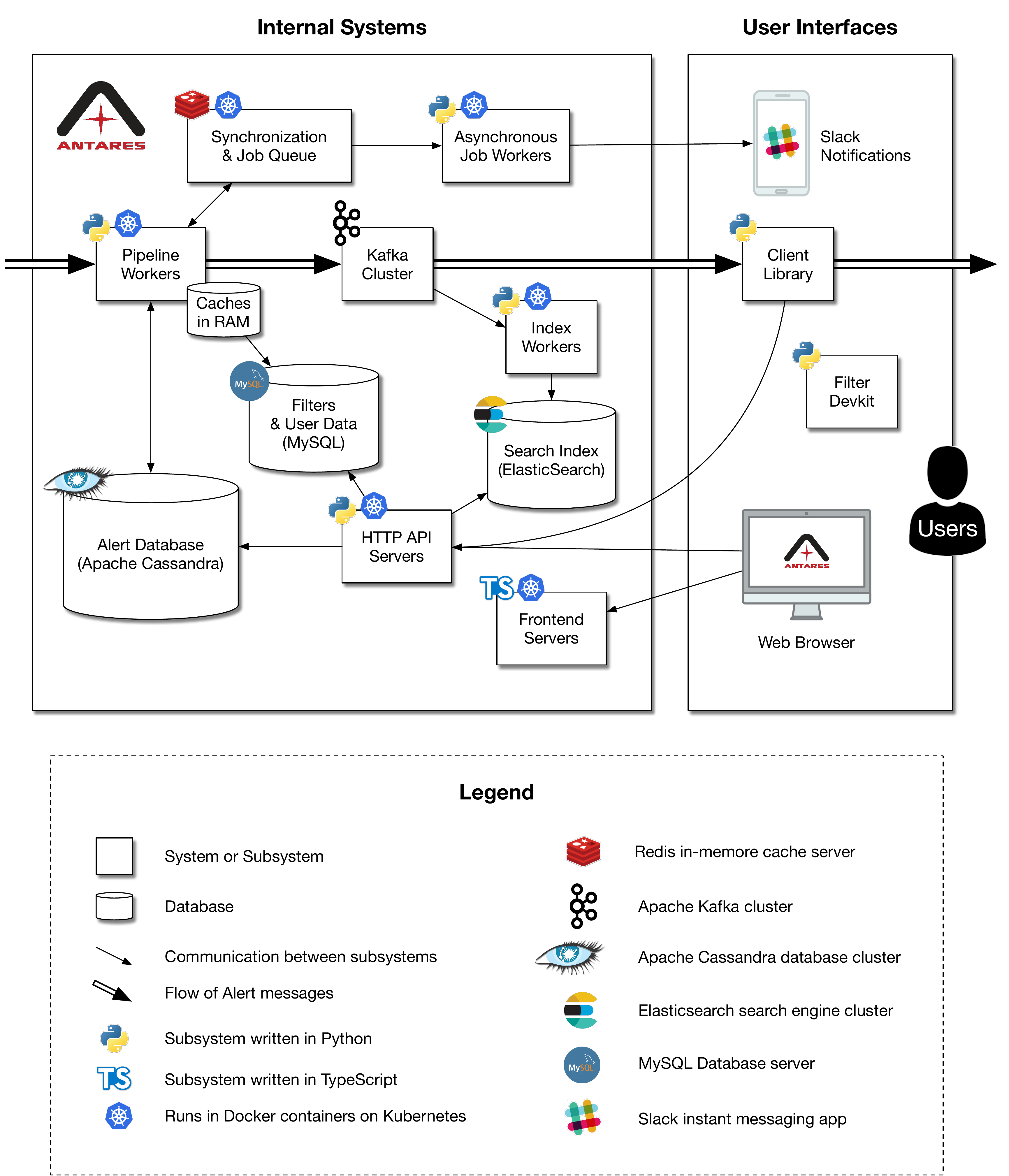}
    }
    \caption{Architecture diagram of ANTARES. Input Alerts enter the system on the left (from ZTF, LSST, etc) and output Alerts exit on the right (to downstream systems, TOMs, etc). Many subsystems represent clusters of processes on multiple machines. Subsystems marked with the blue Kubernetes logo run on a shared pool of hardware resources in a Kubernetes cluster. Subsystems not marked with the Kubernetes logo run on dedicated hardware.}
    \label{fig:architecture}
\end{figure}

We show the architecture of the system in Figure \ref{fig:architecture}. This illustrates the connections between components and the flow of data through the system.

ANTARES processes alerts concurrently in multiple instances of a
program called the ``Pipeline Worker.''  We run one instance for each
partition in the Kafka topic from which they receive data.  Each
worker processes alerts in the order they were received and uses
Kafka's ``commit'' mechanism to track progress and remain resilient to
system failures. The alerts are processed according to the configured
science workload and the results are stored in the ANTARES database as
well as distributed to the community.  The Pipeline Workers load
Filter code from the MySQL
database\footnote{\url{https://www.mysql.com/}} upon startup. Loci,
Alerts, and their annotations are stored in the Alert Database, which
is implemented with Apache
Cassandra.\footnote{\url{https://cassandra.apache.org/}} The Alert
Database is the single source of truth for ANTARES data, although its
content is also indexed separately by the Search Engine. The Alert
Database schema is designed for fast access by the Pipeline Workers
and thus does not support complex queries (see Section
\ref{sec:database}). The Pipeline Workers perform Locus associations
and execute filter code on the Loci (see Section
\ref{sec:pipeline}). After each Locus is processed, updated
information about the Locus is sent to the Search Engine (implemented
using Elasticsearch\footnote{\url{https://www.elastic.co/}}) via Kafka
and the Index Worker microservice. The Search Engine indexes all Loci
and makes them queryable by the Application Programming Interface
(API). The API service is a Representational State Transfer (REST)
Hypertext Transfer Protocol (HTTP) API that is used by the Portal web
site and by the Client Library. The API uses the Search Engine to
perform searches for Loci, and loads the complete data for each Locus
from the Alert Database. The Client Library and the Portal interface
with users, providing visibility into the database.

The Pipeline Workers also produce two other forms of output, Slack notifications and output Kafka Alerts. Slack messages are sent using an asynchronous cluster of job workers that process tasks in a job queue stored in the Redis\footnote{\url{https://redis.io/}} in-memory cache system. The workers implement automatic retry and exponential back off in the event that Slack is not responding. This system is designed to facilitate other forms of notification in the future such as web hooks, emails, or Short Message Service (SMS) text messaging as needed. In ANTARES 1.0, only Slack notifications are implemented.

Also shown in the architecture diagram is the Filter Devkit, described in Section \ref{sec:devkit}.

\subsection{Alert Pipeline}\label{sec:pipeline}

\begin{figure}[p]
    \makebox[\linewidth]{
        \includegraphics[width=0.94\textwidth]{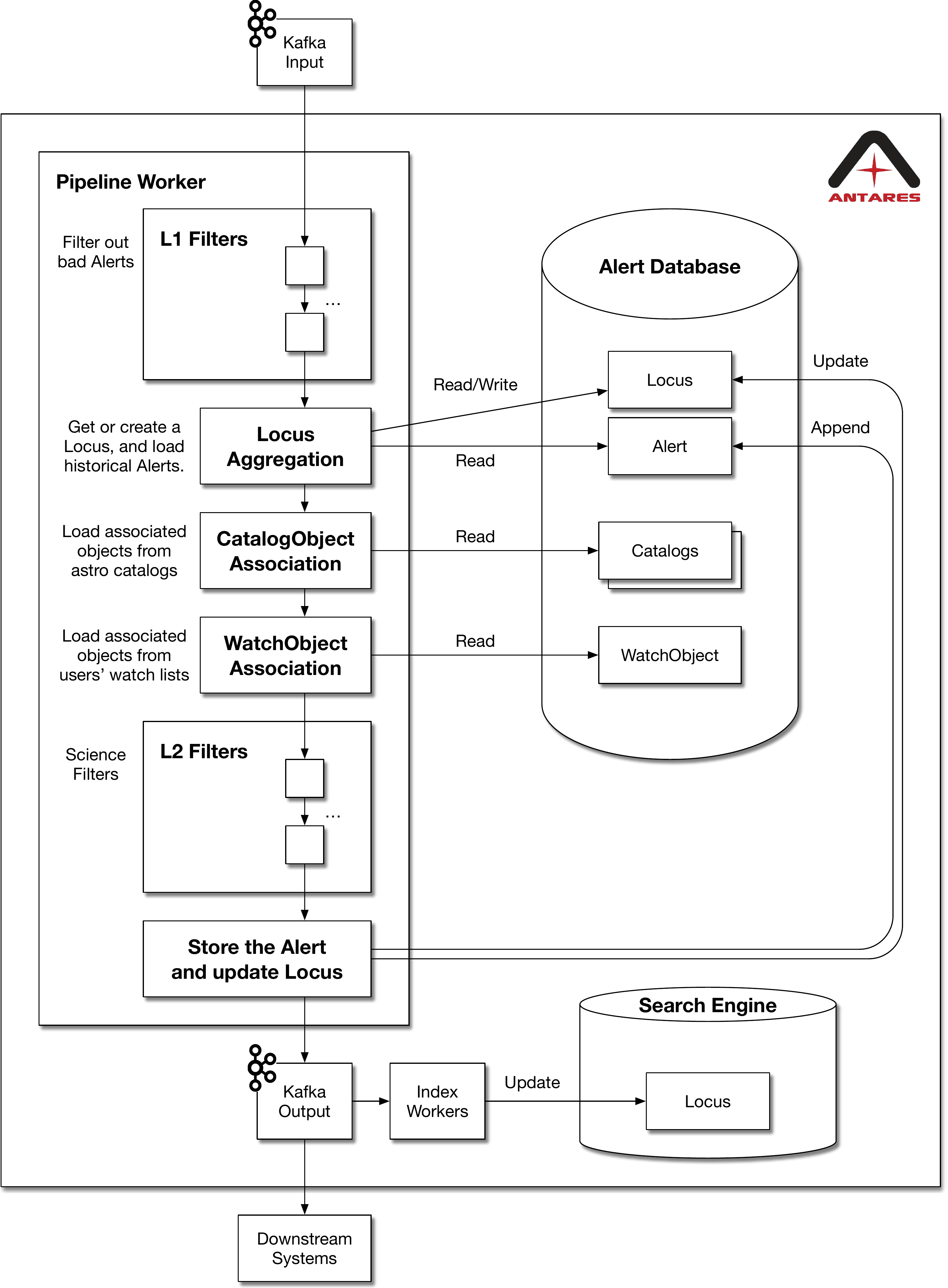}
    }
    \caption{Logical structure of the Alert Pipeline Worker. Input Alert streams are consumed at the top of the diagram and output streams are produced at the bottom.}
    \label{fig:pipeline}
\end{figure}

The internal structure of the Pipeline Workers is shown in Figure \ref{fig:pipeline}. Alerts enter at the top and outputs are produced at the bottom. The first stage of processing is the L1 Filters, that determine whether to run the rest of the pipeline on a given Alert, or whether to simply store it in the database and perform no other processing. This capability can be used to ignore Alerts that are determined to be false detections or of poor quality by some metric. The L1 Filter capability is optional, depending on ANTARES configuration.

The next step is Locus Aggregation, in which the incoming Alert is assigned to a Locus object. This process uses a 1\arcsec\ radius cone search. That is, each incoming Alert is associated with the nearest Locus within 1\arcsec\ of the Alert, with precedence given to
survey-provided associations. If no such Locus exists, a new Locus is created there and is assigned a unique ``locus\_id.'' If there is a pre-existing Locus, then all historical Alerts associated with it are loaded from the database. From this point onward in the pipeline, the object being processed is a Locus object, not an Alert object.

The next stages of our pipeline are CatalogObject Association and WatchObject Association, in which CatalogObjects and WatchObjects are loaded for the given Locus. This is the first step in annotating the Locus with value-added data. This is described in detail in Section \ref{sec:annotation}.

With all known data loaded for the Locus, the pipeline executes the L2 Filters or ``Science Filters.'' These are Filters developed by the ANTARES science staff \citep[e.g.,][]{soraisam20} or submitted by the community.  Filters are described in Section \ref{sec:filters}. They implement science use cases such as classification and other decision making.

Finally, the new Alert is written to the database and the Locus is updated with whatever new properties and associations it has acquired. Output Alerts are broadcast to downstream systems using Kafka and the Search Engine is updated with the new Locus information.

\subsection{Fault Tolerance}

ANTARES runs Filter code submitted by the community, which is difficult to rigorously test. Therefore, ANTARES expects and handles filter crashes. Filters are immediately disabled if they crash and a notification is sent to the author via Slack. The notification contains a ``crash\_log\_id'' and the ``locus\_id'' of the Locus that was being processed, allowing the author to inspect the error and replicate the fault using the Devkit (see Section \ref{sec:devkit}). The author may then improve the Filter, test it on actual data, and submit a new version. Other than temporary downtime of the Filter, there are no negative consequences to this occurring. From experience we have found that this capability is essential when developing new Filters.

Every component in the main ANTARES data pathway is a distributed system, allowing continuous operations in the event of a hardware failure of one machine at a time. Simultaneous failure of multiple machines may require a hands-on systems-administration response.

ANTARES' fault tolerance capabilities benefit from the properties of Kubernetes and Kafka. Specifically, Alerts that are received are not ``committed'' (i.e., marked as received) in Kafka until their processing is complete. Therefore, when a Pipeline Worker crashes in an unexpected way, it can pick up where it left off when it reboots. Kubernetes automatically restarts containers that fail.

Each system in the main data pathway is idempotent. This means that if the same Alert is received more than once, the final data in the Alert is correct. For example, duplicate Alerts are detected and skipped.

\section{Annotation}\label{sec:annotation}

In addition to the base data included in the original alert, ANTARES attaches data and associations to Alerts and Loci. This ``value-added'' data we call Annotations. Annotations have three purposes: to provide richer input data for Filters, to allow Filters to store computed properties on Loci and Alerts, and to allow searchability of the database by meaningful criteria. ANTARES includes several types of annotations, including Locus properties, Alert properties, Locus Tags, astronomical catalog object associations, and associations with watched objects.

Catalog associations are performed by first modeling each catalog object as circular regions of sky of a particular radius. Each Locus is annotated with all catalog objects whose circular region the Locus falls within. In the case of extended objects, the radius of each object region is taken from the catalog data. In the case of point sources, the object is given a default value of 1\farcs

The catalog search algorithm uses a hierarchical triangular mesh \citep[HTM,][]{kunszt01} ID lookup-table (HTMLUT). The HTMLUT allows for efficient and scalable object associations. Each catalog object is entered into the HTMLUT multiple times, once for each HTM trixel ID that the catalog object's circular region intersects. Because HTM trixel regions are triangular and not circular, the region for each object in the HTMLUT is larger than the circular region representing the catalog object. Therefore, false positive associations occur when using the HTMLUT alone. To resolve this, the position and radius of each object are checked after loading their data, and false positives are removed.

The HTMLUT is multi-level, meaning that it supports HTM IDs at multiple levels of tesselation. Each catalog object is represented at a level of tesselation calculated from its radius. The algorithm is tuned to select an HTM tesselation level such that each object intersects 2.5 HTM trixels on average. This level provides an approximate balance between minimizing the number of entries in the HTMLUT and minimizing the number of false positives that have to be loaded, checked, and removed.

The HTMLUT scheme can represent object regions of any size and shape, given an algorithm to represent the region as a set of HTM trixels at various tesselation levels. In ANTARES 1.0, circular regions have been implemented. Ellipses or other shapes can be implemented in the future. 

The exact complement of catalogs will evolve as new catalogs are made available or old ones are updated, but LSST data will be incorporated as catalogs are published.  The current set of catalogs in ANTARES is listed in Table \ref{table:catalogs}, many derived from catsHTM \citep{catshtm}. Several catalogs in ANTARES 1.0 make use of variable radii, including the Third Reference Catalogue of Bright Galaxies (RC3), the Revised Catalog of GALEX Ultraviolet Sources (GALEX), and the The Two Micron All Sky Survey (2MASS) extended source catalogs. In the case of RC3, we adopted a search radius associated with the apparent major isophotal diameter. We are using D25, measured at the surface brightness level $\mu_{B}$ = 25.0 B-mag per square arcsecond. For GALEX, we adopted a search radius associated with the Kron radius in the NUV. In practice, the Kron radius is expressed in the form of NUV\_KRON\_RADIUS * NUV\_A\_WORLD in the GALEX catalog \citep{galex}. We thus adopt this form to define our search radius for GALEX. For the 2MASS extended source catalog, we use the semi-major axis (in arc seconds) of a fiducial ellipse at isophote K=20mag/arcsec$^2$, taken from the catalog data field ``r\_k20fe.''

Catalogs used by ANTARES are stored locally within the Cassandra database. This allows our HTMLUT system to be used and  eliminates the latency and reliability issues associated with querying to external databases hundreds of times per second.

\begin{deluxetable*}{ll}
\tablecaption{External Catalogs Used in ANTARES\label{table:catalogs}}
\tablewidth{0pt}
\tablehead{
\colhead{Catalog} & \colhead{Reference}
}
\startdata
The Two Micron All Sky Survey       & \citet{2MASS} \\
AllWISE Data Release                            & \citet{allwise} \\
ASAS-SN Catalog of Variable Stars & \citet{asassn14, asassn18, asassn19, asassn19b} \\
Second-Generation Guide Star Catalog              & \citet{GSC} \\
\emph{Chandra} Source Catalog              & \citet{chandrasource} \\
Catalina Surveys  & \citet{catalina13a, catalina13b, catalina14,  catalina15, catalina17} \\
Preferred Tidal Disruption Hosts        & \citet{french18} \\
\emph{GAIA} Data Release 2             & \citet{gaiadr2} \\
The NASA/IPAC Extragalactic Database\tablenotemark{a} & \citet{ned} \\
 New York University Value-Added Galaxy Catalog  & \citet{nyu-value} \\
Third Reference Catalogue of Bright Galaxies & \citet{rc3, rc3corwin} \\
Sloan Digital Sky Survey Data Release 12  & \citet{sdssdr12} \\
A Catalogue of Quasars and Active Nuclei: 13th Edition & \citet{veron} \\
The Third XMM-Newton Serendipitous Source Catalogue  & \citet{xmm-newton} \\
Revised Catalog of GALEX Ultraviolet Sources & \citet{galex} \\
\enddata
\tablenotetext{a}{The NASA/IPAC Extragalactic Database (NED) is funded by the National Aeronautics and Space Administration and operated by the California Institute of Technology.} 
\end{deluxetable*}

\section{Filters}\label{sec:filters}

The Filter is the primary way in which users of ANTARES can winnow a stream of time-domain alerts. By evaluating data, both from the alert and annotations, algorithms implemented in filters can identify objects of interest and eliminate irrelevant objects, producing a subset of alerts that astronomers can then pursue with other observational resources.  ANTARES Filters are snippets of Python \citep{vanrossum09} code that process incoming Alerts. Filters can be built in to the ANTARES codebase, or can be submitted by the community using the ANTARES web Portal. Filters are executed in a sequence called the Filter Pipeline (see Figure \ref{fig:pipeline}). The Pipeline runs on each Locus when a new Alert is received on that Locus.  Filters can use data files such as statistical models, neural networks, lookup-tables, etc. These repositories represent distillations of astronomical knowledge derived from larger samples \citep[e.g.,][]{soraisam20}.  These larger samples are what we call a Touchstone for ANTARES, but the Touchstone itself is external to the Pipeline described here.  It is a separate system that enables the design, training, and deployment of filters.

\subsection{Filter Inputs and Outputs}

Filters have access to a variety of information sources in ANTARES at runtime, including the Locus object and its properties and Tags, all catalog objects associated with the Locus, all Alerts associated with the Locus and their properties, and recent LIGO/Virgo detection reports. Timeseries data such as Alert properties over time are available as ``Astropy.TimeSeries'' objects \citep{astropy13} or as Pandas dataframes \citep{pandas10}. Other data are represented as Python data structures.

Based on the input data, Filters may take several actions, including setting properties on the Locus, setting properties on the newly arrived Alert (but not on historical Alerts), adding Tags to the Locus, or halting the pipeline and thus preventing subsequent Filters from running. This last ability is restricted to prevent misuse. Properties may be of type float64, string, or long integer.

\subsection{Filter Structure}

Filters are Python classes that inherit from class ``antares.devkit.Filter'' and implement, at minimum, a method \texttt{run(self, locus).} A simple filter is the HelloWorld filter, shown below. The HelloWorld filter declares that it may produce a Tag called ``hello\_world'' and gives the Tag a text description. Then, in the \texttt{run} method, the filter adds the tag to the locus. There is no conditional logic around adding the tag, so this tag will be added to every locus that the Filter sees.

\begin{verbatim}
import antares.devkit as dk

class HelloWorld(dk.Filter):
    OUTPUT_TAGS = [
        {
            'name': 'hello_world',
            'description': 'This tag is added to EVERY Locus.',
        },
    ]

    def run(self, locus):
        locus.tag('hello_world')
\end{verbatim}

The HelloWorld filter is a trivial example to demonstrate the basic format of a filter.  Appendix \ref{high snr filter} contains a more complex, real-world example.  There, we present a high signal-to-noise ratio filter that explicitly declares all inputs and outputs, performs initial setup, executes a simple computation, conditionally adds a Tag depending on input data, and handles a potential error case.

\section{Outputs}

The output of ANTARES takes three forms.  These are our searchable database, Slack notifications through our Slack workspace,\footnote{\url{https://antares-noao.slack.com}} and Kafka streams. The database is searchable using the ANTARES Portal and can also be queried using the HTTP API either by users or by autonomous systems. Slack notifications are used to inform users and teams in real time of events. This includes newly tagged Loci, new Alerts on tagged Loci, and hits to WatchLists.  ANTARES output Kafka streams are intended to be consumed by downstream systems. Streams are produced by tagging Loci. Each stream is configurable to include a union or intersection of one or more Tags. For example, given a Filter that tags some Loci ``extragalactic'' and another Filter that tags some Loci ``sn1a\_candidate,'' a stream could be configured to include all Alerts whose Loci have both tags. A system could then connect to that stream and receive all such alerts from ANTARES. Connections to the HTTP API and Kafka streams are facilitated by the ANTARES Client Library, discussed in Section \ref{sec:client}.

\section{Provenance Tracking}\label{sec:provenance}

We considered two primary use cases for tracking of provenance data associated with processing of alerts by ANTARES.  First, reproducing offline the state of the ANTARES system at a time in the past, and, second, viewing how filter decision making changed over time on a Locus. An example of the first use case would be if an astronomer wished to inspect an Alert that was processed by ANTARES in the past.  This individual could, based on the provenance data, run a copy of ANTARES with the same configuration as ran in production and reproduce the decisions made while evaluating that alert. As an example of the second use case, given a Filter that produces a classification prediction, a user might wish to view how their Filter's classification of a given Locus changed as each new Alert was received and processed.

To address the first use case, ANTARES logs all of its configuration and state on startup. When each ANTARES process boots up, it records information about the version of the system, the version of all dependencies, catalogs, filters, etc. The full content of the Provenance Log is shown in Table \ref{tab:provenance}. This information is stored as a Provenance Log object in the database, indexed by a provenance\_id. Each incoming Alert is annotated with the provenance\_id of the ANTARES process that received it. Thus, the state of the system that processed a given Alert can be retrieved and, if necessary, reproduced offline. To simplify the state of ANTARES, the system does not add/remove Filters while running. Instead, the system reboots automatically every 24 hours and updates its filter set and configuration at that time only. If necessary, the system can be manually rebooted more frequently with no negative effect on operations.

\begin{deluxetable*}{ll}
\tablecaption{ANTARES Provenance Log Content\label{tab:provenance}}
\tablehead{
\colhead{Provenance Label} & \colhead{Description}
}
\startdata
    antares\_version   & ANTARES package version \\
    config            & ANTARES system configuration \\
    filters           & Active Filters with version numbers \\
    catalogs          & Active Catalogs \\
    python\_packages   & Python package versions \\
 \enddata
\end{deluxetable*}

To address the second provenance use case, ANTARES allows Filters to record time-series of variables. This is implemented using Alert properties. Alert properties can be written to the current (i.e., new) Alert under consideration by a Filter and are immutable thereafter. For example, a Filter could exist that predicts the probability that a Locus is an RR Lyrae variable star. Each time the Filter runs it could store that probability as an Alert property on the new Alert. Then, the Filter's author can view this value over time using the Client library or the Devkit.

\section{Database Design}\label{sec:database}

The data model described above is implemented using two databases, the alert database and the search index. These databases are implemented using Cassandra and Elasticsearch, respectively. Cassandra stores all data in a high-performance and scalable manner, but is indexed only by Locus ID and sky coordinate (represented by HTM IDs). Elasticsearch provides search indexes over values such as Locus properties, Tags, and catalog matches. To reduce capacity requirements, Elasticsearch stores only Locus data, not Alert data.

\subsection{Apache Cassandra -- ANTARES Alert Database}

Cassandra was chosen for its horizontal scalability and performance. The final size of the ANTARES data set is estimated at between 100TB and 2PB. The estimate varies depending on many factors such as the final specification of the LSST alert packet contents, and the number of annotations that Filters produce. Taking the upper limit of 2PB, Cassandra handles volumes of this size in many organizations. Apple famously operates the largest publicly known cluster that, as of 2015, stored 10PB of data on 75,000 nodes\footnote{\url{https://cassandra.apache.org/}}. Cassandra supports the Apache Spark\footnote{\url{https://dl.acm.org/doi/10.1145/2934664}} distributed cluster-computing framework, which opens the possibility of developing batch-processing jobs to run over the entire ANTARES data set in the future. ANTARES uses a Cassandra replication factor of 3 to ensure durability of data.

ANTARES uses custom tooling to monitor and operate our Cassandra cluster, that contains 6 nodes at the time of the ANTARES 1.0 release. These tools report realtime performance and status metrics, gather logs and monitor the health of cluster, and Slack notifications in the event of problems. The tools also allow the scheduling and triggering of Cassandra's built in anti-entropy repair feature that we run weekly. Another custom tool, called ``my2cass,'' allows object catalogs stored in MySQL to be automatically migrated into Cassandra. This entails inspecting the MySQL table schema, creating an equivalent table in Cassandra, copying data into Cassandra using the ANTARES distributed job queue system, and populating the HTMLUT. These investments in operational capability are essential, as the Cassandra cluster will grow significantly over the 10 years of LSST operations.

\subsection{Elasticsearch -- ANTARES Search Engine}

Our Cassandra database is designed to handle the particular set of
queries described in 4.4. This allows it to handle LSST-scale
throughput but prevents users from writing more complex queries to
search for data they want. In order to allow for flexible searches
over the ANTARES data holdings, we additionally store information
about all of the Loci we've processed in an Elasticsearch
cluster. Elasticsearch is a search engine built around Apache Lucene
that allows our team and users to find Loci and compute aggregate
statistics.  It was chosen for its efficient addition of new indexes
to existing tables, and horizontal scalability. Because new Locus
properties, Tags, and Catalogs will be added continuously over time,
the Search Engine needs to be able to efficiently add new
indexes. ANTARES automatically creates appropriately typed indexes in
Elasticsearch as this occurs. At ZTF-scale, a single node
Elasticsearch cluster suffices to store the search index. At LSST
scale the Elasticsearch will need to be expanded out to multiple
nodes.

\section{Performance}\label{sec:performance}

Over the 10 years of LSST operations, the lengths of the light curves of variable objects will increase linearly from zero to hundreds of data points per Locus. Because ANTARES allows Filters to process the complete Locus history data each time an Alert is received, the load on the system will increase linearly over time. This is true both in computational load and read load on the database. Experience with pre-Cassandra designs of ANTARES running on ZTF data have taught us to plan ahead. In the first few months of ZTF data, a single MySQL node could easily act as the Alert Database. After more than a year of ZTF data, this was untenable. A distributed database was required. This linearly increasing load demands that long-term scaling plans be made in advance. No static system configuration that works at year 1 is guaranteed to work at year 10, or even at year 3.

Alert streams have the useful property that each Alert can be processed independently of other nearby Alerts, as long as they are not close enough together that they may belong to the same object (1\arcsec). ANTARES makes use of this property by processing data in parallel. The default configuration in our load tests has been 10 ANTARES processes, each with 10 threads. The number of processes and threads per process can be increased at any time, given sufficient hardware. Likewise, the database needs to scale over time. This is a matter of both capacity and performance. Response times must be low enough that ANTARES keeps up with the incoming data. 

There is one barrier to arbitrarily scaling the system that can be addressed when needed. Kafka topics from ZTF (and from LSST) are internally split into multiple partitions. Kafka distributes messages approximately evenly between partitions. Kafka topic partitions allows a consumer (technically, a ``consumer group'') to pull messages from a Kafka topic using multiple processes simultaneously. The number of processes can only be as large as the number of partitions. If, in the future, this becomes a bottleneck, ANTARES may simply re-partition the Kafka stream into more partitions by mirroring it into a local Kafka cluster with more partitions. 

Although alert processing occurs in parallel, there are two process coordination tasks that occur at the beginning of the processing of each alert. First, a mutex-like\footnote{A mutex, or mutual exclusion object, allows multiple program threads to share the same resource, but not to use the resource simultaneously.} locking mechanism prevents two Alerts on the same Locus from being processed at the same time. This mechanism is called the HTM Region Lock. Implemented using Redis and Lua\footnote{\url{http://www.lua.org/}} scripts, the region lock allows processes to lock access to sets of HTM trixels in an atomic, thread safe manner. This is a bottleneck. However, the load on this system is constant over time and does not increase as light curves lengthen. In testing, the region lock introduces 10ms to 100ms of latency into the pipeline when processing Alerts at a test rate of 10 alerts/second. This small latency cost is acceptable to prevent race conditions. If this latency grows to unacceptable levels at LSST alert rates, then we will spread the load over multiple Redis instances.

Second, in the event of creating a new Locus object, a unique ``locus\_id'' is generated using a counter stored in memory in Redis. Earlier ANTARES designs used random GUIDs such as ``1c165092-a540-11ea-94be-324c811e1e2.'' By using an integer counter and converting the integer to a compressed string format, Locus IDs can instead take a human-readable form such as ``ANT2020so7ia''. This system was inspired by the ZTF Object ID scheme, (e.g., ``ZTF18aabejqj'') and produces shorter, more recognizable identifiers than UUIDs. As with the HTM Region Lock, the performance is acceptable. The system recovers gracefully from Redis failures.

\subsection{Scalability Properties}

We conjecture that ANTARES scales in proportion to three variables: the rate of incoming alerts, R, the number of historical alerts stored in the database, N (the integral of R over time), and the total computational complexity of filters, F. Note that N is somewhat related to the average length of all light curves in the database, which grow over time as the survey operates. We further conjecture that the number of worker processes and the number of Cassandra database nodes both must scale in proportion to a function of R and N, while the CPU usage of the system scales in proportion to a function of all three variables R, N, and F.

Note that the value of N increases linearly with time as it integrates R. Therefore even with constant R, the load on ANTARES increases over time.

An estimate of the alert rate, R, can be made based on LSST's estimated value of 10 million alerts per night and a typical night lasting 10 hours. This gives an average alert rate of about 280 alerts per second over the course of the night. Over shorter time intervals the rate will very greatly over the course of a night, with spikes of much higher rates. Since alerts are queued and buffered in Kafka, ANTARES does not need to meet the maximum spike throughout rate, but it does need to easily handle the average rate over the course of each night.

\subsection{ZTF-Scale Load Test Results}

\begin{figure}
    \makebox[\linewidth]{
        \includegraphics[width=0.8\textwidth]{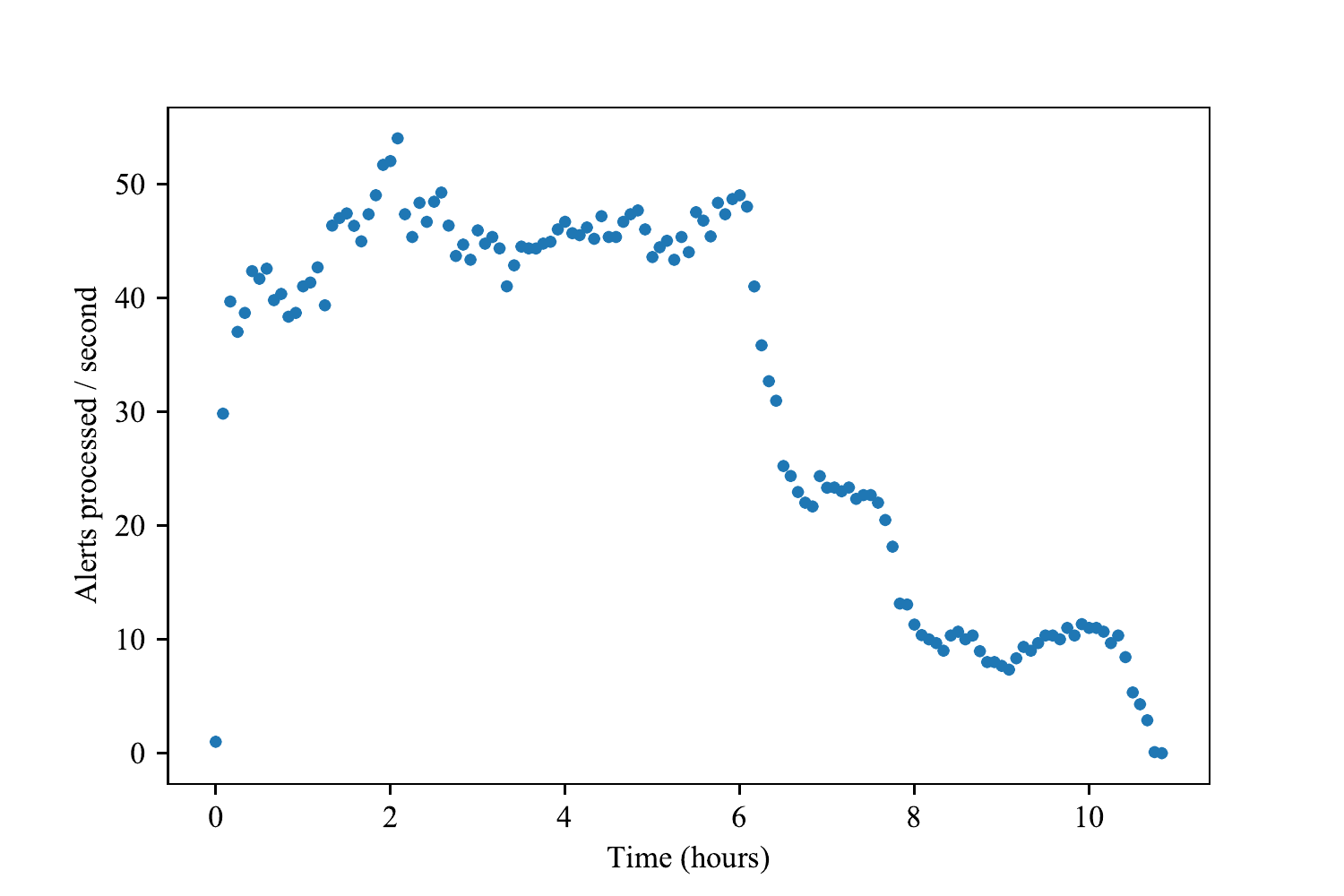}
    }
    \caption{Alert throughput over time as ANTARES processes 1.23 million ZTF alerts in approximately 11 hours. The data set was created by merging eight consecutive nights of ZTF data into a singe input stream for ANTARES.}
    \label{fig:throughput}
\end{figure}

We present the results of a load test of ANTARES using 1.23 million alerts taken from eight consecutive nights of ZTF data. The alerts were loaded into a single Kafka topic (stream) with 20 partitions on a Kafka cluster of 10 nodes running on a cloud service provider. ANTARES was deployed on-site at NOIRLab and was configured to connect to this stream with 20 worker processes. Each process had 10 threads for a total of 200 threads. The system under test was the ANTARES infrastructure and database, not a particular combination of filters. A single filter was enabled with an average execution time less than 1 millisecond. Alert processing rate over time is shown in Figure \ref{fig:throughput}. A steady-state throughput rate of approximately 45 alerts/second is visible for the first 6 hours. The rate then drops step wise until all alerts have finished being processed at approximately t=11 hours.

The step wise decrease in throughput rate towards the end of the processing run has been observed during multiple tests, and appears to be scale-invariant with respect to the number of alerts in the run. We conjecture that this phenomenon is an artifact of Apache Kafka. Kafka divides each stream into multiple parallel sub-streams called partitions. A group of processes that together consume a single topic is called a ``consumer group.'' Each individual consumer process in the consumer group may consume from multiple partitions, but each partition may only be consumed by one consumer process at a time. Some partitions finish being processed before others, and their consumers then sit idle while the slower partitions finish. We suspect that this causes a ``tail'' effect on plots of throughput over time, and is consistent with the step wise nature of the tail towards the end of a processing run. We will continue to investigate this.

We have observed that steady-state throughput rate of ANTARES is dependent on which Kafka input cluster is used and on the number of partitions in the topic. We measure CPU, memory, and local network well below full utilization. This leads us to speculate that ANTARES capability with current hardware is limited by the input stream rather than by the database, local network, or CPU and memory constraints. This may be due to the internet connection to the Kafka cluster, to the number of nodes in the Kafka cluster, or to the number of partitions in the topic. We will investigate this further by using a local Kafka cluster directly connected to the ANTARES local network, and by varying the number of nodes and the number of partitions.

\subsection{Alert Processing Time}

\begin{figure}
    \makebox[\linewidth]{
        \includegraphics[width=0.7\textwidth]{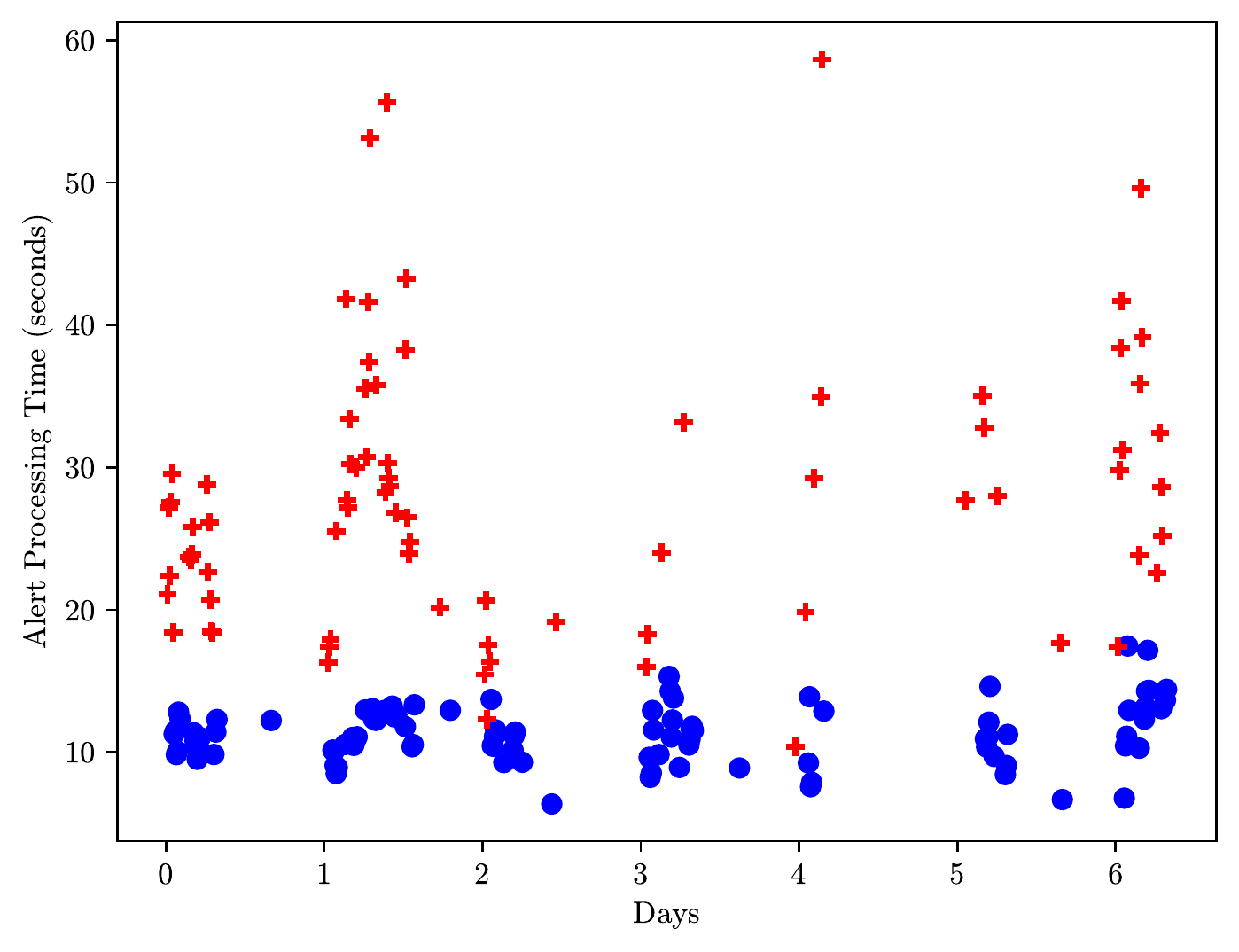}
    }
    \caption{Data from several days of operation showing the time that an alert spends in the ANTARES pipeline.  Blue dots represent median values while red crosses show the maximum time spent in the pipeline for each of the operating windows plotted.}
    \label{fig:alert_process_time}
\end{figure}

Figure \ref{fig:alert_process_time} shows the typical times that ANTARES takes to process alerts. We define this as the time that an alert spends in the ANTARES pipeline from beginning to end. These data were taken from 7 days of real-time processing of ZTF alerts. Because ANTARES is multi-process and multi-threaded, many alerts are in some stage of the pipeline at any given moment. The total throughput has been tested at 45 alerts/second as described above, but each alert spends multiple seconds in the pipeline. In Figure \ref{fig:alert_process_time}, we show that the median alert processing time is around 11 seconds and the maximum is typically between 30 and 60 seconds.

\subsection{Opportunities for Performance Increases}

As discussed above, every subsystem of ANTARES is scalable including the Cassandra database and the Alert Pipeline. LSST throughput rates will be about 10x higher than the test result presented above. Also, LSST operations will demand continuous scaling over time because the size of the alert database will grow linearly. We expect to achieve the required scale by expanding the ANTARES server clusters. In addition to simply adding servers, we will continue to pursue performance improvements to the system. Cassandra is highly tunable and we will continue to investigate how best to configure it for our particular query paths.

In the event that technology or funding limits our ability to scale the system, we have contingency options. One option is to use Level 1 Filters more aggressively to filter out artifacts and other non-astrophysical alerts. Another is to discard the majority of input alert properties and keep only an essential subset. Depending on the content of the LSST alert packet, this could significantly decrease our database capacity needs while preserving our science use cases. Another option is to store all data, but only process a subset of Loci in real-time according to some filtering. Other data could be stored and deferred for daytime processing (although the day only adds a factor of two, as night will fall again). We consider these options to be undesirable and we do not plan to implement them. We expect to achieve the required scale through adding nodes and by tuning Cassandra and the ANTARES pipeline.

\section{User Interface}

The ANTARES user interface has three components, the Web Portal, the API and companion Client library, and the Filter Devkit.

\subsection{Web Portal}\label{sec:portal}

\begin{figure}
  \includegraphics[width=\linewidth]{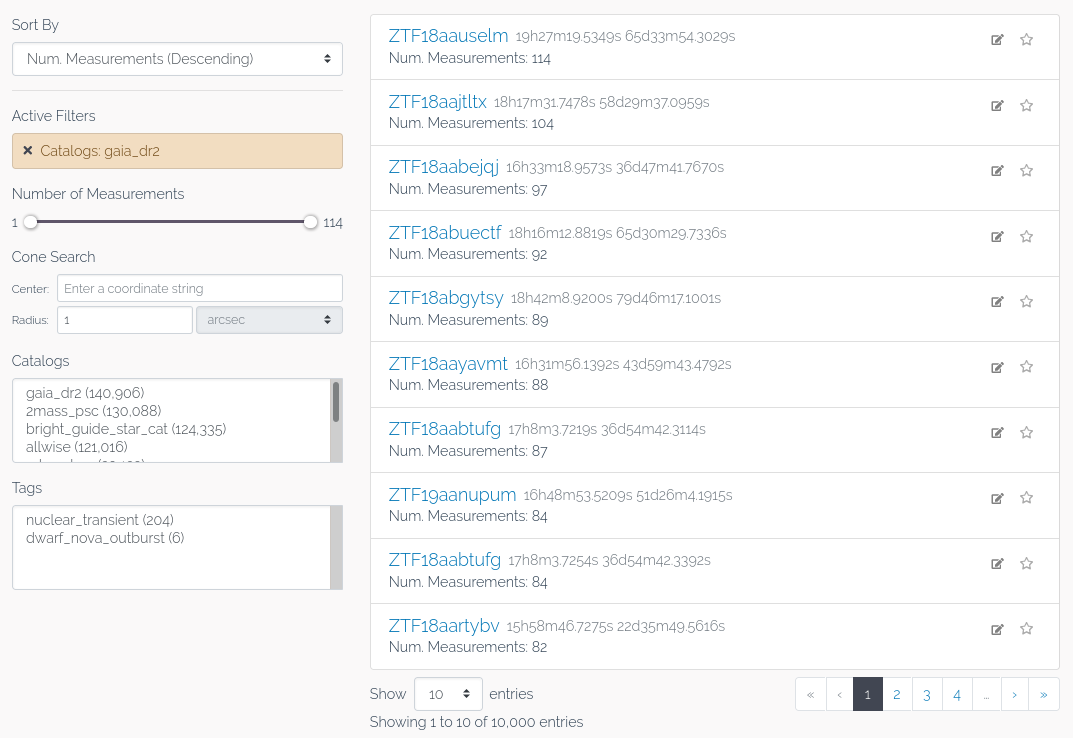}
  \caption{Web-based search interface for the ANTARES Portal. Users can interactively refine search result sets along a number of different dimensions, including properties of the object (its location and the number of measurements there), catalog associations, and annotations added to the locus by filters (tags). Users are also able to bookmark and make private annotations on loci.}
  \label{fig:portal_explore}
\end{figure}

\begin{figure}
  \includegraphics[width=\linewidth]{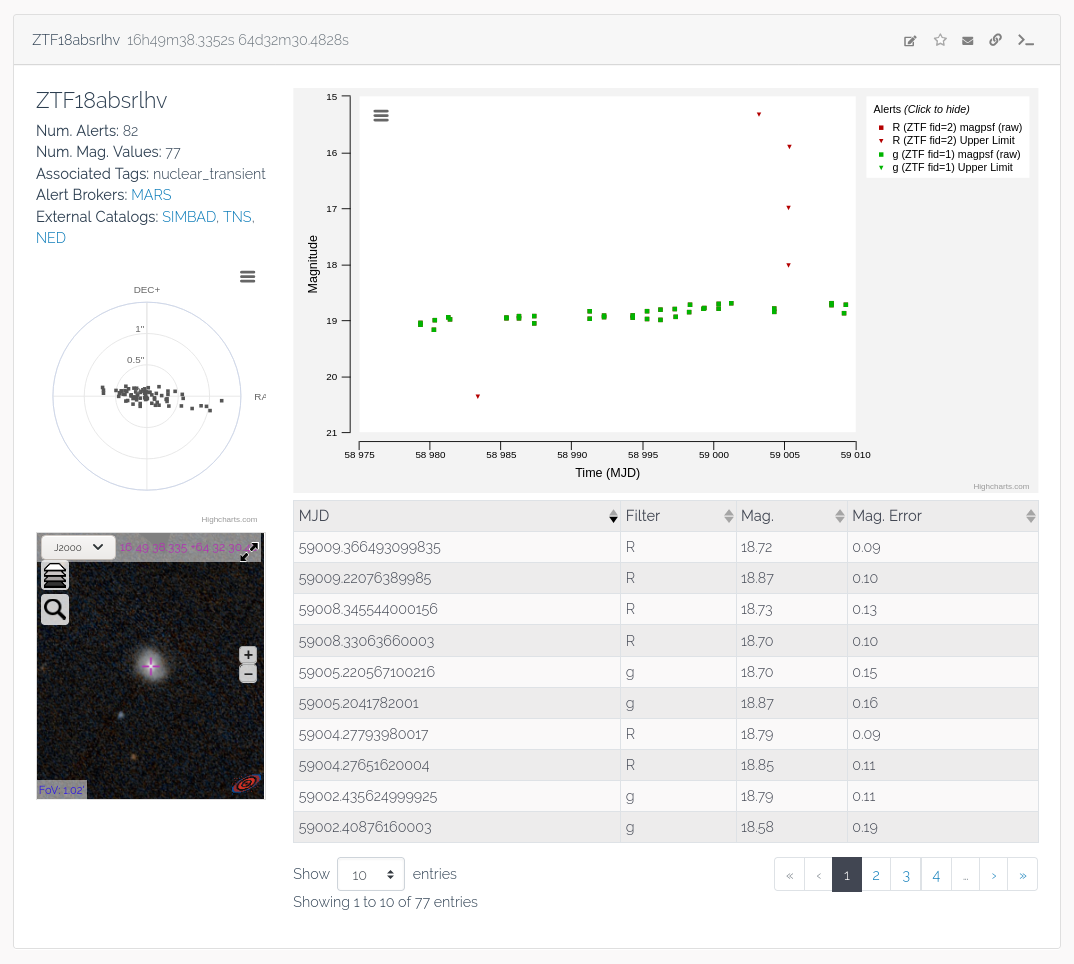}
  \caption{Viewing the details of a locus on the ANTARES Portal.  Users can see a plot of the light curve of the object, a table of observed values, links to external catalogs, the positions of detections relative to the locus center, and a thumbnail view of the region powered by Aladdin Lite \citep{bonnarel00, boch14}.}
  \label{fig:portal_locus}
\end{figure}

The ``Portal" is a client-side web application that lets scientists submit filters and visually explore our database. The homepage (\url{https://antares.noirlab.edu/}) shows the most recent data that the system has processed, aggregate statistics about the ANTARES data holdings (see Figure \ref{fig:portal_explore}), and details about particular loci (see Figure \ref{fig:portal_locus}). Because manually constructing Elasticsearch queries (for the locus search engine) is difficult, we allow users to interactively build search queries for loci that meet certain criteria and then to export them for use in the ANTARES client (see Section \ref{sec:client}). The Portal also provides an administrative interface for submitting, approving, and managing filters and watch lists in the pipeline.

\subsection{Application Programming Interface and Client Library}\label{sec:client}

ANTARES allows for interactive and programmatic data access through an HTTP API. This API exposes ``resources," views into our database that generally align  with the data models described in Section \ref{sec:datamodel}. For example, the request \texttt{GET /loci} will receive as response a collection of ``locus" resources that each have a light curve, a set of locus properties, etc. We chose to build the API in accordance with version 1.0 of the JavaScript Object Notation:API (JSON:API) specification.\footnote{\url{https://jsonapi.org/format/1.0/}} This decision was motivated by a number of considerations. A JSON:API supports partial fetching of resources (e.g., fetch just the light curve of a locus) and the ability to include related resources in a request (e.g., fetch a locus and its most recent alert). It also allows for hypermedia-driven interactions that improve discoverability of our data and simplify maintaining backwards-compatibility as we continue development work. In short, it allowed us to build an API that supports a rich range of queries, minimizes network traffic, and leverages tools from an existing ecosystem of software for producing and consuming JSON:API-conformant data. 

We also provide a user-friendly Python library, the ``ANTARES Client," that supports scientists in consuming streaming data through Kafka streams and in analyses of the ANTARES data holdings through the HTTP API. It also includes a command-line interface that enables common tasks such as running a daemon to save a local copy of all the data from a particular stream. A copy of this client library is hosted on the Python Package Index\footnote{\url{https://pypi.org/project/antares-client/}} so it can be installed with the Python package manager ``pip." Currently, the interfaces provided in the client access resources through the API that don't require the user to be authenticated. We plan to expand these in the future so that users can programmatically access data such as objects of interest that they have bookmarked. Consuming data from Kafka streams requires credentials that the ANTARES team provides upon request. Documentation for installing and using the client library is provided as a standalone website.\footnote{\url{https://noao.gitlab.io/antares/client/}}

The Client library was used to discover the first R Coronae Borealis star from the ZTF public survey \citep{Lee20}. Candidates were preselected using color-color cuts, and their long term light curves were loaded from the alert database. Upon further inspection of the light curves, we discovered ZTF18abhjrcf showing large (greater than 5 magnitudes) brightness variation over an extended period of time (about 100 days). This is similar to the behavior of known R Corona Borealis star. Further spectroscopic follow-up with the Las Cumbres Observatory telescope has confirmed its R Corona Borealis classification.

\subsection{Filter Devkit}\label{sec:devkit}

The Devkit allows filter authors to develop and test their filter code using real data from the ANTARES database. It is designed to be used in the NOIRLab's DataLab\footnote{https://datalab.noao.edu/} Jupyter environment, where it has direct read-only access to the ANTARES databases. Any user may sign up for Datalab and use the Devkit. The Devkit allows users to fetch Locus data from the database and run filter code on this data in a Python environment that mimics the ANTARES production system, as demonstrated in the following example.

\begin{verbatim}
import antares.devkit as dk
dk.init()  # Initialize the Devkit system

# Define a filter:
class HelloWorld(dk.Filter):
    def run(self, locus):
        pass

# Run the filter on a random Locus from the database:
dk.run_filter(HelloWorld)

# Run the filter on a specific Locus from the database:
dk.run_filter(HelloWorld, locus_id=...)
\end{verbatim}

Users can also construct custom Locus data to test their filters, either by modifying real data or constructing it completely from scratch. The Devkit documentation describes how to do this. Some filters require access to data files such as statistical models, neural networks, lookup-tables, etc., as described in Section \ref{sec:filters}. These data files can be uploaded into ANTARES using the Devkit. Full documentation for the Devkit is provided as a standalone website. The Portal's FAQ page provides links to all documentation.

\section{ANTARES in the Era of LSST}

The data instrument we have described here is already functional at the scale of alerts produced by ZTF.  The step to LSST scale will be at least an order of magnitude greater, if not two.  We are confident that the design of the system will scale to match LSST, with only database tuning and hardware expansion necessary to accommodate the larger number of alerts.  Specifically we expect the Cassandra cluster to require significant expansion both for capacity and for throughput.  In addition to providing real-time reactivity to alerts, ANTARES will provide a database of all LSST alerts over the operational period, and is designed to allow future work on batch processing of this data set.  ANTARES will be a general-use broker for the US and world community as it takes advantage of all the time-domain science opportunities that LSST will provide.

\acknowledgments

The ANTARES team would like to thank the following individuals for their support and advice in the development of this software system: Tim Axelrod, Robert Blum, Todd Boroson, Glenn Eychaner, Mike Fitzpatrick, Michael Fox, Steve Howell, Tim Jenness, Tod Lauer, Nirav Merchant, Catherine Merrill, Robert Nikutta, Knut Olsen, Stephen Ridgway, Robert Seaman, Claire Taylor, Adam Thornton, Jackson Toeniskoetter,  Alistair Walker, and John Wregglesworth.

The ANTARES team gratefully acknowledges financial support from the
National Science Foundation through a cooperative agreement with the
Association of Universities for Research in Astronomy (AURA) for the
operation of the NSF's National Optical-Infrared Astronomy Research
Laboratory, through an NSF INSPIRE grant to the University of Arizona
(CISE AST-1344024, PI: R. Snodgrass), and through a grant from the
Heising-Simons Foundation (2018-0909, PI: T. Matheson).

%

\vspace{5mm}


\software{
    Python\footnote{\url{http://python.org}} \citep{vanrossum09},
    Astropy\footnote{\url{http://www.astropy.org}} \citep{astropy13},
    Pandas\footnote{\url{https://pandas.pydata.org/}} \citep{pandas10, reback2020pandas},
    Kafka,\footnote{\url{https://kafka.apache.org/}}
    Apache Cassandra,\footnote{\url{https://cassandra.apache.org/}}
    Kubernetes,\footnote{\url{https://kubernetes.io/}}
    MySQL,\footnote{\url{https://www.mysql.com/}}
    Slack,\footnote{\url{https://slack.com/}}
    Elasticsearch,\footnote{\url{https://www.elastic.co/}}
    Redis,\footnote{\url{https://redis.io/}}
    Lua,\footnote{\url{http://www.lua.org/}}
    JSON\footnote{\url{https://www.json.org/json-en.html}}
}



\appendix
\section{Example Filter}
\label{high snr filter}

This filter selects alerts that have a high signal-to-noise ratio (SNR), based on data intrinsic to the alert itself.  Downstream filters can use this information to operate only on those alerts whose detection is more secure.

\begin{verbatim}
import antares.devkit as dk

class HighSNR(dk.Filter):
    ERROR_SLACK_CHANNEL = ""  # Put author's Slack user ID or channel here.
    # ^ This will notify the author if the Filter ever crashes.
    
    # The Filter declares that it does not require any LocusProperties.
    REQUIRED_LOCUS_PROPERTIES = []
    # The Filter declares that it requires the following AlertProperties:
    REQUIRED_ALERT_PROPERTIES = [
        'ant_passband',
        'ant_magerr',
    ]

    # The Filter declares that it does not produce output LocusProperties.
    OUTPUT_LOCUS_PROPERTIES = []
    # The Filter declares that it does not produce output AlertProperties.
    OUTPUT_ALERT_PROPERTIES = []
    # The Filter declares that it may produce a Tag called "high_snr"
    # and defines a description for it.
    OUTPUT_TAGS = [
        {
            'name': 'high_snr',
            'description': 'Locus has one or more Alerts with high SNR.',
        },
    ]

    def setup(self):
        """
        Declare constants.

        `setup()` is called once per Filter per thread, at time of system boot.
        Filter performance can be optimized by doing as much work as possible in
        advance in `setup()`.
        eg: loading files, constructing datastructures, etc.
        """
        # The threshold is dependent on the band that is being imaged.
        # These threshold values should flag ~2-3% of alerts.
        self.snr_threshold = {
            'g': 50.0,
            'R': 55.0,
        }

    def run(self, locus):
        """
        If this Alert has a high SNR, then tag the Locus "high_snr".
        """
        passband = locus.alert.properties['ant_passband']  # Get Alert's band
        magerr = locus.alert.properties['ant_magerr']  # Get Alert's mag error

        if passband not in snr_threshold:
            return  # This passband is not supported. Do nothing.
        threshold = self.snr_threshold[passband]  # Get threshold for this band

        snr = 1.0 / magerr  # Compute SNR
        if snr > threshold:
            locus.tag('high_snr')
\end{verbatim}

\bibliography{antares.bib}{}
\bibliographystyle{aasjournal}


\end{CJK*}
\end{document}